# Token-Budgeted Escalation for Financial Document Question Answering: Cost Is Predictable, Benefit Is the Bottleneck

Junru Zhu
*Independent Researcher*
Seattle, USA
junru.zhuu@gmail.com

Yixin Yang
*Independent Researcher*
New York, USA
vicki_yang@chiefgroup.com.hk

Xiaoqing Ding
*University of Chicago*
Chicago, USA
alexading@uchicago.edu

Ruoyu Qi
*Independent Researcher*
Charlotte, USA
dxfreedom94@gmail.com

***Abstract*—Retrieval-augmented generation systems can route difficult queries to deeper context, but batch deployments must allocate a shared token budget across calls whose costs vary by query. We formulate selective escalation as finite-batch allocation for financial document question answering. Each of 150 FinanceBench questions first receives a top-1 retrieval answer. Predictors estimate the adjudication-quality gain and token cost of an optional top-5 call, and the allocator prioritizes calls by predicted gain per token. At the nominal 10% budget, gain-per-token allocation improves adjudication quality over gain-only ranking by 0.034 (95% document-bootstrap confidence interval [0.001, 0.072]) while using 46.6% fewer total tokens than one-pass top-5 retrieval. Additional-call cost is accurately predictable (R-squared 0.93), whereas beneficial escalation remains difficult to rank (area under the receiver operating characteristic curve 0.60). These results show that heterogeneous cost is actionable under tight constraints, while progress across the full budget frontier depends on stronger query-specific benefit estimates.**

## I. INTRODUCTION

Retrieval-augmented generation (RAG) improves document question answering by supplying external evidence [1], but deeper retrieval also increases prompt length, latency, and inference cost. FLARE and DRAGIN trigger retrieval from model uncertainty or evolving information needs [2–3]. Adaptive-RAG and Self-RAG condition retrieval on question complexity or model feedback [4–5], while document-routing and recoverability-aware methods use query and provisional-answer signals [6–7]. IntrAgent combines section ranking with iterative reading for content-grounded literature retrieval [8]. Recent work studies uncertainty, utility, and calibrated correctness under retrieval budgets [9–11]. RAGScope uses lightweight evidence gates to triage generated answers for review or stronger verification [12]. We study batch deployments with a shared token budget and heterogeneous incremental retrieval costs.

This heterogeneity changes the allocation problem. Ranking queries only by predicted benefit treats two equally promising escalations alike even when one requires far more tokens. We therefore separate benefit prediction from cost prediction and prioritize escalation by predicted gain per additional token. The resulting policy allocates computation according to opportunity cost: under a fixed batch budget, an escalation is valuable only relative to the other gains that the same tokens could purchase. We also evaluate a predicted-cost 0/1 knapsack to test whether explicit combinatorial allocation improves on this greedy rule.

We study all 150 public FinanceBench questions from 84 source documents [13]. Retrieval is restricted to evidence pages from the supplied source document, and the generator is fixed. Every question first receives a top-1 retrieval answer. Using only information available after this route, cross-fitted models estimate the benefit and additional-call token cost of issuing a separate top-5 call. We compare gain-only ranking, gain-per-token allocation, and predicted-cost knapsack policies under a common predicted budget and report the tokens each policy actually consumes. A matched-realized-token sensitivity analysis aligns the two learned rankings to within 0.3 tokens per question.

At a budget equal to 10% of escalating the full batch, gain-per-token allocation achieves adjudication quality 0.524, compared with 0.490 for gain-only ranking. The paired difference is 0.034 (95% document-bootstrap confidence interval (CI) [0.001, 0.072]). The policy uses 1340 tokens per question, 46.6% fewer than one-pass top-5 retrieval. Escalation costs are heterogeneous, with a coefficient of variation (CV) of 0.56, but accurately predictable ($R^2$=0.93), whereas beneficial escalation is difficult to identify (area under the receiver operating characteristic curve (AUROC) 0.60). This asymmetry supports a focused conclusion: cost-aware prioritization improves allocation at a tight budget, while further frontier improvements depend principally on better benefit prediction.

Our contributions are:

- a finite-batch selective-escalation formulation in which queries have heterogeneous additional-call costs under a shared token constraint;

- a two-stage allocation pipeline that predicts benefit and cost after top-1 retrieval and prioritizes candidates by expected gain per token;
- a document-grouped evaluation demonstrating a tight-budget advantage for cost-aware allocation and identifying benefit prediction as the main empirical bottleneck.

## II. Token-Budgeted Escalation

### A. Allocation objective

Selective escalation allocates a finite additional-call budget across a batch after every question has completed the cheap route. For question $i$, the cheap route produces quality $q^c_i$ using $t^c_i$ tokens. A separate expensive call produces quality $q^e_i$ using $t^e_i$ tokens. Because this call is issued after the cheap answer, its additional deployment cost is $c_i=t^e_i$, and its quality gain is $g_i=q^e_i-q^c_i$.

The cheap-route cost is common to all allocation policies. Maximizing final batch quality therefore reduces to selecting escalations with the largest aggregate predicted gain. Given an additional-call budget $B$, an allocation $x_i \in \{0,1\}$ maximizes

$$\sum_i x_i \hat{g}_i \text{ subject to } \sum_i x_i \hat{c}_i \le B. \quad (1)$$

where $\hat{g}_i$ and $\hat{c}_i$ are available before the expensive generation. The final quality is $q^c_i+x_i g_i$. Decisions use predicted costs, while reported expenditure uses the realized $\sum_i x_i c_i$. This distinction measures policy quality at the tokens actually consumed rather than at the nominal predicted budget.

### B. Two-stage prediction

The allocation task uses top-1 term frequency–inverse document frequency (TF–IDF) page retrieval as the cheap route and a separate top-5 call as the expensive route. Direct, top-2, and one-pass top-5 routes are retained as evaluation references but are not allocation actions. Fig. 1 summarizes the decision sequence.

The decision-time representation combines question length, numerical structure, and question type; retrieval scores, margin, entropy, coverage, redundancy, and context size; top-1 answer confidence and token counts; and the size and page count of the candidate top-5 context. Candidate-context statistics require page ranking but not the expensive generation. The top-5 answer, its output length, realized top-5 token use, reference answer, and correctness labels are excluded.

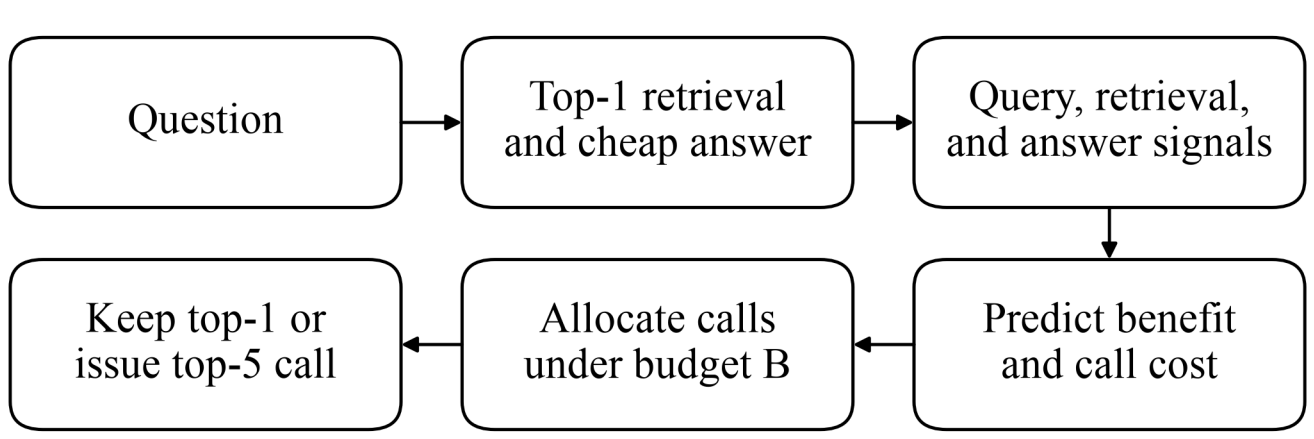

Fig. 1. Two-stage selective escalation. The allocator observes the top-1 answer and pre-generation retrieval signals, predicts the benefit and token cost of an additional top-5 call, and selects calls under a shared budget.

The benefit target is the continuous difference between top-5 and top-1 adjudication quality. Its predictor averages a standardized ridge regressor ($\alpha$=10) and a histogram gradient-boosting regressor with learning rate 0.05, 150 iterations, at most seven leaf nodes, and $L_2$ regularization 1.0. A standardized ridge regressor ($\alpha$=5) predicts additional-call tokens, with predictions lower-bounded at one token. We generate out-of-fold predictions with five-fold cross-validation grouped by source document, so a document never contributes to both training and held-out router observations. The gradient-boosting fit is repeated with seeds 11, 22, and 33, and policy results are aggregated across these runs.

### C. Budgeted allocation

Gain-only allocation sorts candidates by $\hat{g}_i$ and greedily accepts each call whose predicted cost fits the remaining budget. Gain-per-token allocation instead sorts by $\max(\hat{g}_i,0)/\max(\hat{c}_i,1)$ and applies the same feasibility rule. It therefore prefers expected quality improvements that consume fewer additional tokens. The predicted-cost knapsack directly solves the objective above after rounding costs upward to 10-token units and clipping negative predicted gains to zero.

Random order, low top-1 confidence, and low retrieval-score margin use the same greedy feasibility rule as allocation baselines. An oracle-cost knapsack combines predicted gains with realized costs to isolate cost-prediction error, while the oracle gain–cost policy uses both realized gains and costs as an upper bound. All deployable policies select with predicted costs; their comparisons use realized additional-call tokens.

The gain-per-token order is optimal for the fractional relaxation, in which an item can be purchased partially, but is a heuristic when calls are indivisible. The predicted-cost knapsack tests whether exact optimization of the discretized 0/1 objective improves on that order. Both policies still depend on the same estimated gains and costs. Optimization can rearrange candidates under the budget, but it cannot recover benefit information absent from the predicted values. Cost normalization can nevertheless change which calls fit, especially when costs are more predictable than benefits.

## III. Experiments and Analysis

### A. Evaluation protocol

FinanceBench pairs financial questions with source filings [13]. We use all 150 public questions from 84 documents and the 168 unique evidence pages distributed with the dataset. Retrieval is conditioned on the supplied source-document identity, while page selection remains automatic. Our hybrid retriever uses TF–IDF [14] and combines word unigrams and bigrams with character n-grams of lengths 3–5 using weights 0.7 and 0.3.

Amazon Nova Micro [15] generates the direct, top-1, top-2, and top-5 routes at temperature zero with at most 384 output tokens. The completed run contains 600 cached generations, covering four routes for every question. All routes use the same instruction, generator, and decoding settings; retrieved context is the route-dependent input. Input and output token counts and latency come from the inference response. Dollar estimates use the dated price snapshot in the configuration, but all efficiency conclusions use tokens.

Evaluation has two primary outcomes. Motivated by evaluation using a large language model (LLM) as a judge and its documented position effects [16], a route-blind Nova Lite judge receives all four candidate answers with their route identities hidden. Candidate labels are independently permuted across three passes. Majority correctness defines route accuracy, and the mean [0,1] score defines adjudication quality. Deterministic token F1 and numerical agreement are retained as diagnostics. The three passes per question produce 450 joint adjudications. Because generation precedes policy evaluation, every allocator selects from the same logged route outcomes rather than triggering a new, policy-specific model run.

We evaluate nominal budget fractions of 0, 10, 20, 30, 40, 50, 75, and 100%. For deployable policies, a fraction β gives predicted budget $B=\beta\sum_i\hat{c}_i$; oracle policies use the corresponding realized-cost total. Tables and figures report realized total tokens per question, including the completed top-1 call. Statistical comparisons use 2,000 paired bootstrap resamples over source documents. A difference is statistically distinguishable when its 95% interval excludes zero.

Two grouping choices keep related observations together. Cross-fitting assigns all questions from a source document to the same fold, preventing questions from one filing from contributing to both router training and held-out prediction. Bootstrap resampling also operates on documents rather than individual questions, preserving within-document dependence in the uncertainty intervals. The bootstrap holds the out-of-fold predictions and selected sets fixed; it quantifies document-sampling uncertainty conditional on the fitted router and does not include router-refitting variability. Selection uses predicted cost, whereas tables report realized cost, so policy feasibility and actual resource consumption remain separate quantities throughout the analysis.

### B. *Why optimize continuous adjudication quality?*

Binary correctness changes for only 26 helpful and 14 harmful escalations; the remaining 110 questions provide no correctness difference between top-1 and top-5. The continuous adjudication score supplies a signed training target even when an answer improves or degrades without crossing the majority-correctness threshold. It therefore offers more resolution than a binary help label for learning the router.

We retain accuracy as a separate outcome because a higher judge score need not correspond to a corrected answer. At the 10% budget, quality increases from 0.490 to 0.524, and accuracy increases descriptively from 46.0% to 49.3%. The accuracy difference is 3.3 percentage points, with document-bootstrap interval [0.0, 7.6] percentage points; this interval includes zero. The statistically distinguishable primary comparison therefore uses continuous adjudication quality rather than binary accuracy.

### C. *Does deeper retrieval improve quality?*

Table I establishes the value and cost of retrieved evidence. Direct generation reaches 4.7% accuracy, compared with 46.7% for top-1 retrieval. Increasing depth to top-2 and top-5 raises accuracy to 52.0% and 54.7%, respectively, but also increases token use from 1091 to 1719 and 2508 tokens per question. Top-2 reaches quality 0.529 with substantially less context than top-5, motivating selective rather than uniform deep retrieval. Because escalation executes both the top-1 and top-5 calls, the one-pass top-5 route is an efficiency reference rather than a two-stage policy.

TABLE I. ONE-PASS ROUTE OUTCOMES. QUALITY IS THE MEAN ROUTE-BLIND ADJUDICATION SCORE; TOKENS AND LATENCY ARE PER QUESTION.

| Route | Accuracy (%) | Quality | Tokens/q | Latency (ms) |
|---|---|---|---|---|
| Direct | 4.7 | 0.063 | 138 | 334 |
| Top-1 | 46.7 | 0.490 | 1091 | 402 |
| Top-2 | 52.0 | 0.529 | 1719 | 374 |
| Top-5 | 54.7 | 0.567 | 2508 | 388 |

### D. *Does gain-per-token allocation help under a tight budget?*

At the nominal 10% budget, gain-per-token allocation reaches quality 0.524 at 1340 realized tokens per question (Table II). Gain-only ranking reaches 0.490 at slightly greater token use. Their paired quality difference is 0.034, with 95% document-bootstrap interval [0.001, 0.072]. Gain-per-token allocation also uses 46.6% fewer total tokens than one-pass top-5 retrieval. This result supports the central claim at the tightest nonzero operating point: accounting for heterogeneous call cost improves which questions receive deeper retrieval.

We also compare the learned rankings at matched realized token use. Reducing the gain-only nominal budget to 9.817% yields 1340.5 tokens per question, compared with 1340.3 for gain-per-token. At this matched point, quality is 0.492 for gain-only and 0.524 for gain-per-token, a difference of 0.032 with document-bootstrap interval [-0.001, 0.068]. The point estimate persists, but this post hoc matched-cost interval includes zero.

The predicted-cost knapsack improves over gain-only ranking by 0.028, but its interval [-0.003, 0.062] includes zero. Explicit combinatorial optimization is therefore not distinguishable from gain-only ranking in this comparison; the supported advantage belongs to gain-per-token prioritization.

### E. *Why does cost normalization change the selected set?*

The two learned rankings spend nearly the same total additional-call budget but compose it differently. Gain-per-token selects 18 calls with mean realized cost 2077 tokens, whereas gain-only selects 11 calls averaging 3509 tokens. The ratio policy therefore converts predictable cost variation into seven additional opportunities for deeper retrieval rather than simply spending less. The common predicted budget is 37773 tokens; realized spend is 37389 tokens for gain-per-token and 38594 for gain-only, each within 2.2% of the predicted limit.

TABLE II. POLICIES AT THE NOMINAL 10% ADDITIONAL-CALL BUDGET, COMPARED USING REALIZED TOTAL TOKENS.

| Policy | Accuracy (%) | Quality | Tokens/q | Escalate (%) |
|---|---|---|---|---|
| Random | 46.2 | 0.488 | 1335 | 10.0 |
| Margin rule | 46.7 | 0.495 | 1337 | 6.7 |
| Gain only | 46.0 | 0.490 | 1348 | 7.3 |
| Gain/token | 49.3 | 0.524 | 1340 | 12.0 |
| Learned knapsack | 48.7 | 0.517 | 1341 | 12.0 |
| Oracle | 56.0 | 0.584 | 1340 | 11.3 |

TABLE III. ROUTE BEHAVIOR BY QUESTION FAMILY. ACCURACY IS IN PERCENT; COST IS THE MEAN NUMBER OF TOKENS IN THE ADDITIONAL TOP-5 CALL.

| Type | n | Top-1 acc. | Top-5 acc. | Help/Hurt | Cost |
|---|---|---|---|---|---|
| Domain | 50 | 46.0 | 58.0 | 10/4 | 3338 |
| Metrics | 50 | 36.0 | 34.0 | 6/7 | 1616 |
| Novel | 50 | 58.0 | 72.0 | 10/3 | 2569 |

Thus the policy difference comes primarily from allocation composition rather than aggregate budget drift.

The realized outcomes of those selected sets explain the quality difference. 6 of the 18 gain-per-token calls improve correctness, compared with 1 of the 11 gain-only calls; each set contains two harmful escalations, with the remainder neutral. These labels are post hoc diagnostics, not selection inputs, and do not imply that cheaper calls are intrinsically more useful. They show how an expensive candidate with a high but uncertain benefit score can occupy budget that otherwise supports several opportunities. The learned knapsack reaches quality 0.517 at the same 12% escalation rate as gain-per-token, further indicating that a stronger optimizer does not by itself repair noisy benefit values.

### *F. Where does deeper retrieval help?*

The aggregate route averages conceal substantial variation across the three equally sized FinanceBench question families (Table III). Top-5 retrieval raises accuracy from 46.0% to 58.0% on domain-relevant questions and from 58.0% to 72.0% on novel-generated questions. In contrast, metrics-generated accuracy decreases from 36.0% to 34.0%. The help/hurt counts show the same pattern: domain and novel questions have more than twice as many helpful as harmful escalations, whereas metrics-generated questions have six helpful and seven harmful outcomes.

Question family is informative but does not determine the decision. Every family contains both helpful and harmful cases, so a hard type-based routing rule would discard useful within-family variation. Cost also differs across families: domain questions average 3,338 tokens per additional call, compared with 2,569 for novel and 1,616 for metrics questions. A deployable router must therefore estimate query-specific benefit within these broad categories while accounting for their different resource profiles.

Cost quartiles expose a second interaction (Table IV). The most expensive quartile contains 13 helpful and three harmful escalations; its accuracy rises from 44.7% to 71.1%. The cheapest quartile contains one helpful and five harmful cases, and its accuracy falls from 60.5% to 50.0%. Low cost is therefore not a proxy for high value. Gain-per-token must preserve strong benefit evidence for expensive calls while preventing uncertain, high-cost candidates from consuming the batch budget. Cost prediction remains well behaved across the range: quartile-level mean absolute percentage error (MAPE) varies from 8.1% to 12.5%.

TABLE IV. OUTCOMES BY REALIZED ADDITIONAL-CALL COST QUARTILE. EACH QUARTILE CONTAINS 37 OR 38 QUESTIONS; MAPE EVALUATES HELD-OUT COST PREDICTIONS.

| Cost quartile | Tokens | MAPE (%) | Help/Hurt | Acc. top-1→top-5 |
|---|---|---|---|---|
| Q1 (low) | 919 | 8.1 | 1/5 | 60.5→50.0 |
| Q2 | 1716 | 10.2 | 8/4 | 37.8→48.6 |
| Q3 | 2881 | 8.6 | 4/2 | 43.2→48.6 |
| Q4 (high) | 4503 | 12.5 | 13/3 | 44.7→71.1 |

TABLE V. NOMINAL BUDGET ABLATION. DEPLOYABLE POLICIES ALLOCATE USING PREDICTED COSTS, AND TOKENS PER QUESTION REPORT REALIZED USE.

| Budget | Policy | Accuracy (%) | Quality | Tokens/q |
|---|---|---|---|---|
| 10% | Margin | 46.7 | 0.495 | 1337 |
| 10% | Gain only | 46.0 | 0.490 | 1348 |
| 10% | Gain/token | 49.3 | 0.524 | 1340 |
| 10% | Knapsack | 48.7 | 0.517 | 1341 |
| 10% | Oracle | 56.0 | 0.584 | 1340 |
| 20% | Margin | 46.7 | 0.496 | 1605 |
| 20% | Gain only | 48.0 | 0.509 | 1606 |
| 20% | Gain/token | 48.7 | 0.524 | 1600 |
| 20% | Knapsack | 49.3 | 0.526 | 1605 |
| 20% | Oracle | 61.3 | 0.634 | 1584 |
| 50% | Margin | 54.7 | 0.565 | 2338 |
| 50% | Gain only | 52.0 | 0.545 | 2381 |
| 50% | Gain/token | 50.7 | 0.529 | 2373 |
| 50% | Knapsack | 50.7 | 0.529 | 2373 |
| 50% | Oracle | 64.0 | 0.663 | 1919 |

### *G. How does the advantage change across budgets?*

Table V and Fig. 2 show that policy ordering changes with the operating region. At 20%, gain-per-token and knapsack allocation reach quality 0.524 and 0.526 near 1,600 tokens per question, compared with 0.509 for gain-only ranking. At 50%, the retrieval-margin rule reaches quality 0.565, while gain-only, gain-per-token, and knapsack reach 0.545, 0.529, and 0.529. Cost-aware allocation consequently has a supported tight-budget advantage but does not dominate the complete frontier.

The oracle reaches quality 0.634 at 20% and 0.663 at 50%, remaining above all deployable policies. The route outcomes therefore contain useful allocation structure that the learned benefit scores do not recover. Reporting the full set of predeclared budgets characterizes this transition rather than selecting an operating point after observing the results.

Across nonzero budgets, realized spend differs from the predicted limit by at most 2.7%. Gain-only and gain-per-token share 6 calls at 10% (Jaccard 0.26), rising to 0.94 at 75%. Cost normalization therefore changes allocation mainly when resources are scarce; at larger budgets, benefit-ranking error dominates.

### H. What limits allocation quality?

Top-5 escalation helps 26 questions, hurts 14, and leaves 110 unchanged (Fig. 3). Additional-call cost is heterogeneous: realized costs range from 588 to 5498 tokens, with the middle half spanning 1364–3656 tokens and CV 0.56. This variation is predictable. The out-of-fold cost model obtains $R^2$=0.93, mean absolute error 263 tokens, and 9.8% mean absolute percentage error. The continuous benefit model has $R^2$=-0.08 and mean absolute error 0.337. Its score obtains AUROC 0.60 and an area under the precision-recall curve (AUPRC) of 0.22 when used to rank questions whose correctness improves. Cost can therefore be estimated much more reliably than benefit in this experiment.

At the 10% budget, replacing predicted cost with realized cost in the knapsack changes quality from 0.517 to 0.520, whereas the full gain–cost oracle reaches 0.584. Most observed headroom therefore lies in benefit estimation rather than cost estimation. This asymmetry is consistent with the frontier: accurate cost estimates help avoid unusually expensive calls under a tight budget, while benefit-ranking errors increasingly determine allocation as the budget grows.

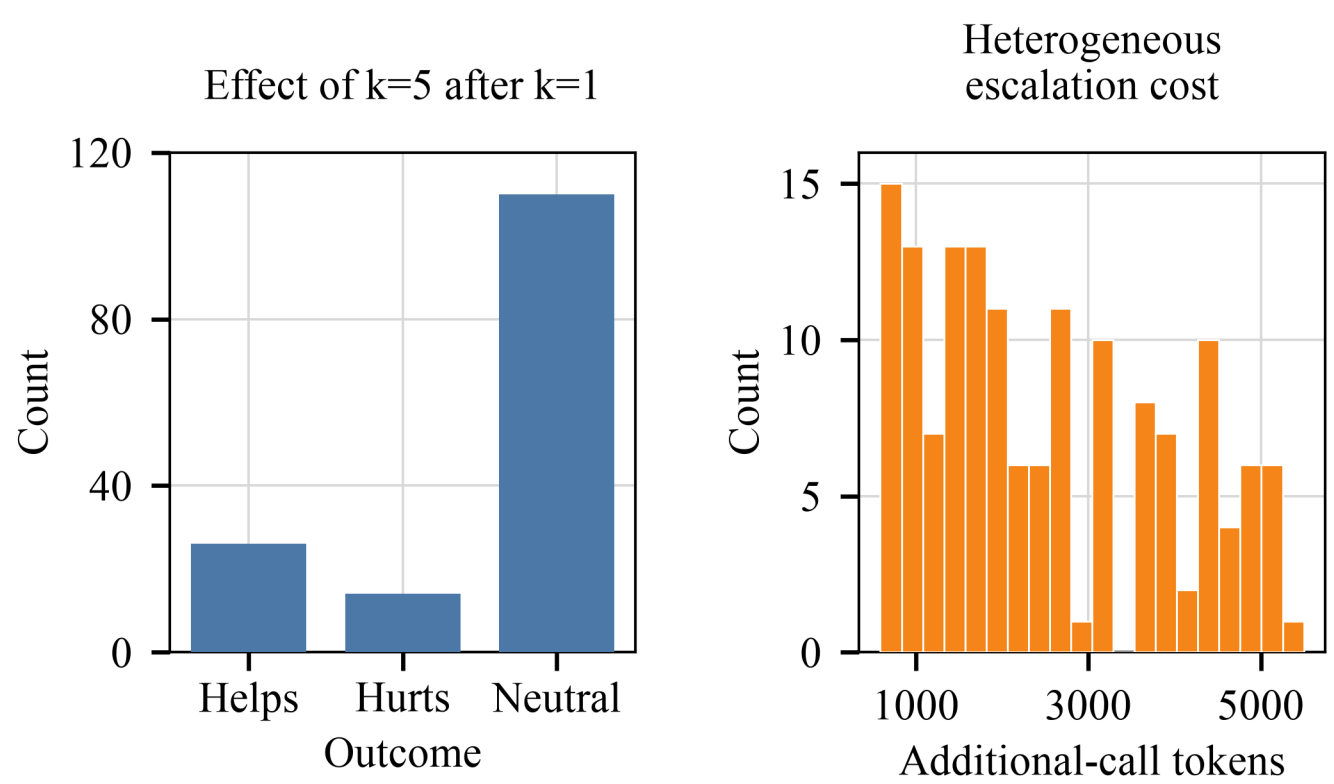

Fig. 3. Observed top-5 versus top-1 outcomes and additional-call token cost. Most questions have neutral correctness change, producing a sparse benefit target despite heterogeneous cost.

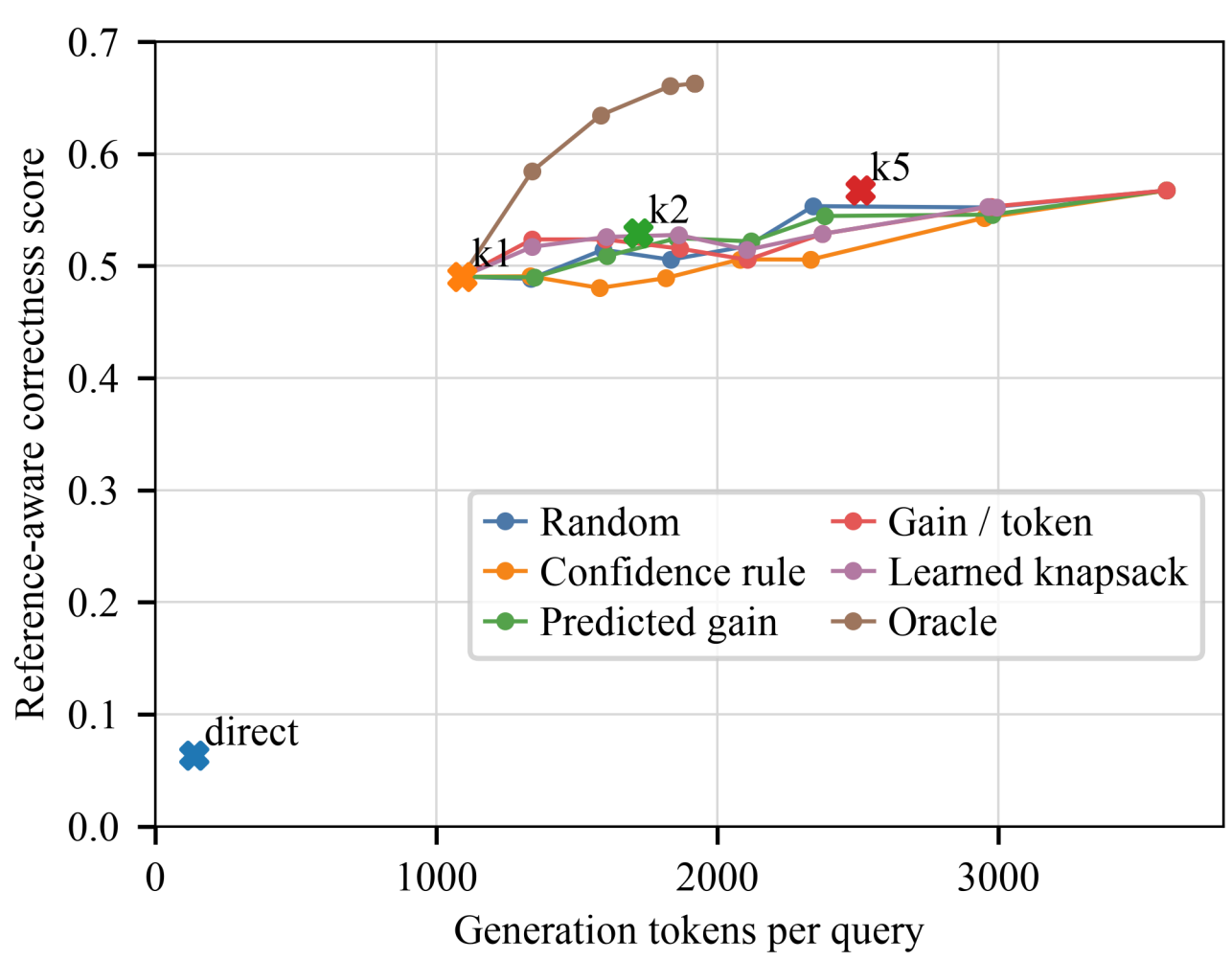

Fig. 2. Adjudication quality versus realized total tokens. Gain-per-token allocation improves the tight-budget region, while policy ordering changes as the budget grows.

### I. Implications for batch deployment

The allocation procedure maps directly to a batch scheduler. After cheap-route calls complete, the scheduler computes decision-time features, predicts one benefit and cost per query, and issues expensive calls only for the selected subset. Gain-per-token requires a sort and a linear feasibility pass, giving O($n$ log $n$) selection time for a batch of $n$ queries. The knapsack uses rounded token units and trades additional computation for exact optimization of its discretized predicted objective. If a service enforces a hard token cap, it can reserve a calibration margin around predicted costs; realized token use should still be logged because predicted feasibility does not guarantee exact billing.

The resource variable can also reflect a different deployment constraint. Tokens are appropriate when prompt volume or inference spend is binding. A latency-constrained service could instead predict incremental latency or optimize a joint resource budget, although the present latency measurements depend on the provider and concurrent load. Likewise, a system that reuses the top-1 context or intermediate computation would define $c_i$ as the remaining incremental work rather than the full top-5 call. Such reuse changes the quality–cost frontier and requires a new cost model, but the finite-batch allocation objective remains unchanged. Iterative LLM workflows also arise in network operations: Xiao et al.'s CORA combines domain-constrained prompts with validation feedback for multi-vendor OpenROADM automation [17].

## IV. Limitations

The evidence covers 150 FinanceBench questions, one generator, and one source-conditioned TF–IDF retriever. Supplying the source-document identity and a curated evidence-page pool removes document discovery, Portable Document Format (PDF) acquisition, optical character recognition (OCR), and cross-document retrieval from the evaluated task. The results therefore establish the tight-budget allocation effect in this controlled financial question answering (QA) setting rather than across domains or model families. The sample contains only 26 helpful escalations, which constrains learning and uncertainty estimation for the sparse benefit target.

Answer quality relies on route-blind Nova Lite adjudication rather than human judgment. Independent candidate permutations reduce route-position effects but do not remove model-judge bias. Token F1 and numerical agreement provide deterministic diagnostics, although neither fully captures the correctness of free-form financial answers. Reported latency also depends on the model provider and concurrent load.

The per-question audit is descriptive: its failure categories overlap and lack causal annotations.

## V. Conclusion

We formulate selective escalation as finite-batch allocation under heterogeneous additional-call token costs. On 150 FinanceBench questions, gain-per-token allocation improves adjudication quality from 0.490 to 0.524 over gain-only ranking at the nominal 10% budget, while using 46.6% fewer tokens than one-pass top-5 retrieval. This advantage does not extend uniformly across larger budgets. Additional-call cost is accurately predictable ($R^2$=0.93), whereas beneficial escalation remains difficult to rank (AUROC 0.60). Cost-aware prioritization can therefore improve tightly constrained financial document QA, while further frontier gains depend on better query-specific benefit estimation.